\documentclass[aps,prc,preprint,showpacs,showkeys,floatfix,nofootinbib]%
{revtex4}
\usepackage{graphicx}
\usepackage{color}

\newcommand{\bmath}[1]{\mbox{\boldmath${#1}$}}
\newcommand{\ri}[1]{{\bf r}_{#1}}

\newcommand{\bp}{{\bf p}}
\newcommand{\qi}[1]{{\bf q}_{#1}}

\newcommand{\bsigma}{\bmath{\sigma}}

\newcommand{\beps}{\bmath{\epsilon}}

\newcommand{\phat}{\bmath{\widehat{\bf p}}}

\newcommand{\rhat}{\bmath{\widehat{\bf r}}}

\newcommand{\expup}[1]{e^{#1}}

\newcommand{\EG}{{\textrm{e.g.}}}
\newcommand{\IE}{{\textrm{i.e.}}}
\newcommand{\EA}{{\textit{et al.}}}

\date{\today}

\begin{document}
\title{Near threshold \boldmath$NN\to d\pi$ reaction in chiral 
perturbation theory}

\author{A. G{\aa}rdestig}\thanks{Corresponding author:
Anders G{\aa}rdestig,
Department of Physics and Astronomy,
Ohio University,
Athens, OH 45701,
U.~S.~A.}\email{anders@phy.ohiou.edu}
\author{D. R. Phillips}\email{phillips@phy.ohiou.edu}
\author{Ch. Elster}\email{elster@phy.ohiou.edu}
\affiliation{Department of Physics and Astronomy, 
Ohio University, Athens, OH 45701}

\begin{abstract}
The near-threshold $np\to d\pi^0$ cross section is calculated in
chiral perturbation theory to next-to-leading order in the expansion
parameter $\sqrt{M m_\pi}/{\Lambda_\chi}$.  At this order irreducible
pion loops contribute to the relevant pion-production operator.  While
their contribution to this operator is finite, considering initial-
and final-state distortions produces a linear divergence in its matrix
elements.  We renormalize this divergence by introducing a
counterterm, whose value we choose in order to reproduce the threshold
$np \rightarrow d \pi^0$ cross section measured at TRIUMF.  The
energy-dependence of this cross section is then predicted in chiral
perturbation theory, being determined by the production of $p$-wave
pions, and also by energy dependence in the amplitude for the
production of $s$-wave pions.  With an appropriate choice of the
counterterm, the chiral prediction for this energy dependence
converges well.
\end{abstract}


\maketitle

\section{Introduction}
Ever since the days of current algebra, soft-pion theorems, which
relate processes involving different numbers of pions, have been used
to understand threshold pion-production reactions~\cite{KR,SSY}.  
These relations originate from the chiral symmetry of the underlying
Lagrangian of quantum chromodynamics (QCD) and have now been
systematized into an effective field theory called chiral perturbation
theory ($\chi$PT)~\cite{We79,ulfreview}.  In $\chi$PT,
there is not only an understanding of these relations between
amplitudes, but also an organizing principle (power counting), which
suggests a hierarchy of contributions to the amplitudes and thus makes
it possible to assign a theoretical error to a calculation.  
$\chi$PT can be regarded as an extension of and improvement on current 
algebra and it has found many applications. 
Significant progress has been made in the pion-pion and pion-nucleon sectors 
(for a review of the latter see~\cite{ulfreview}) and for nucleon-nucleon
scattering~\cite{Weinberg,bira,evgeny}.  Here we will apply it to threshold
pion production in few-nucleon systems.

The application of chiral perturbation theory to threshold pion-production
reactions has met with mixed success. In the charged-pion production
reactions the leading pion rescattering mechanism involves the
Weinberg-Tomozawa (WT) $\pi N$ amplitude. Considering this, plus
certain higher-order effects, produces moderate agreement between
theory and data cross sections~\cite{dRMvK}. 
Approaches to these reactions based on meson-exchange models are also 
quite successful~\cite{CHpolpi,Teresa}.

Meanwhile, both meson-exchange models and $\chi$PT initially struggled
to describe the $pp\to pp\pi^0$ cross section measured close to
threshold. Data from the Indiana University Cyclotron
Facility~\cite{Meyer} and the The Svedberg Laboratory in Uppsala,
Sweden~\cite{TSL} were five times larger than phenomenological
calculations~\cite{Miller}.  Two early proposals to remedy this
deficiency were short-range interactions~\cite{LeeRiska,Chucketal} and
$\pi N$ amplitudes with a particular off-shell dependence~\cite{Oset}.
In $\chi$PT these effects are both higher order, and they are related
to each other through field redefinitions~\cite{Cohen}.  This
sensitivity to higher-order effects is partly a consequence of the
particular isospin structure of the $pp\to pp\pi^0$ reaction, which
forbids the pion rescattering via the WT term that is the leading
effect in the charged-pion channels.  (In $\chi$PT there is also a
partial cancellation between higher-order one-body and rescattering
terms, which is not present in meson-exchange model
treatments~\cite{Cohen,Pa96,USC1,USC2}.)  But even at a formal level,
it quickly became apparent that the conventional $\chi$PT expansion of
amplitudes in $\frac{m_\pi}{M}$ (with $M$ and $m_\pi$ the nucleon and
pion masses) does not apply to pion production in few-nucleon systems,
since the initial nucleons have a large relative momentum
$P=\sqrt{Mm_\pi+\frac{m_\pi^2}{4}}\sim 360$~MeV/c.  Because of this an
expansion in $P/M=\sqrt{m_\pi/M}\sim0.4$ was adopted~\cite{Cohen}.
But the convergence of this expansion is slow, and it is necessary to
include several poorly constrained higher-order mechanisms in order to
achieve a good description of the data, \EG, the short-range
interactions and/or off-shell amplitudes mentioned above~\cite{vK96}.  These
effects are included in meson-exchange-model calculations of $pp\to
pp\pi^0$, but even so such calculations struggle to describe spin
observables~\cite{CHpolpi,IUCFpolpp}.  The present situation, with a
detailed review of this interesting field, is given in
Ref.~\cite{CHreview}.  As that review points out, in spite of its
difficulties, the $\chi$PT expansion in powers of $\sqrt{m_\pi/M}$
does predict the correct hierarchy of mechanisms for the production of
pions in a relative $s$-wave.  It also successfully explains $p$-wave
pion production~\cite{CHp}.

The $np\to d\pi^0$ reaction is of particular and current interest.
The total cross section was measured some time ago at TRIUMF~\cite{Hutcheon}, 
and fitted to the expression
\begin{equation}
  \sigma(np\to d\pi^0) = \frac12(\alpha\eta+\beta\eta^3),
\end{equation}
where $\eta\equiv q/m_\pi$ is the reduced c.m.\ pion momentum and
Ref.~\cite{Hutcheon} gives the values $\alpha=184\pm5~\mu$b and
$\beta=781\pm79~\mu$b for what they call the reduced (pion) $s$- and
$p$-wave cross sections (but see also below).  The factor $1/2$
results from a comparison with the (Coulomb corrected) isospin partner
$pp\to d\pi^+$.  The $\pi^+d\to pp$ data (applying detailed balance)
are in agreement with $np\to d\pi^0$~\cite{Ritchie}.  However, more
recent data on the $pp\to d\pi^+$ reaction gives
$\alpha=208\pm5$~$\mu$b, $\beta=1220\pm100$~$\mu$b~\cite{Heimberg} and
$\alpha=205\pm9$~$\mu$b, $\beta=791\pm79$~$\mu$b~\cite{Drochner},
indicating a possible isospin violation.  A model
calculation~\cite{Chuck} managed to reproduce the data of
Ref.~\cite{Hutcheon}, but it was subsequently pointed out that
Ref.~\cite{Chuck} overestimated contributions from short-range
pion-production mechanisms~\cite{Jouni}.  More recently the
charge-symmetry-breaking (CSB) forward-backward asymmetry was measured to be 
$A_{\rm fb}=[17.2\pm8({\rm stat})\pm5.5({\rm sys})]
\times10^{-4}$~\cite{Allena}.  
The asymmetry is an interference between charge independent (CI) and CSB 
$s$- and $p$-waves. 
Ignoring relative phases it can be expressed as
\begin{equation}
  A_{\rm fb}(np\to d\pi^0) \sim 
  \frac{\sqrt{\alpha\widetilde\beta}\eta^2+\sqrt{\widetilde\alpha\beta}\eta^2}
       {\alpha\eta+\beta\eta^3}\approx 
       \sqrt{\frac{\widetilde\beta}{\alpha}}\eta+
       \frac{\sqrt{\widetilde\alpha\beta}}{\alpha}\eta,
\label{eq:Afb}
\end{equation}
where $\widetilde\alpha$ and $\widetilde\beta$ are the CSB equivalents of 
$\alpha$ and $\beta$.  
It is believed that this asymmetry, together with the recent
$5\sigma$ result of the $dd\to\alpha\pi^0$ experiment~\cite{IUCFCSB}, will help
constrain some of the lesser known, but important, CSB parameters in
the chiral Lagrangian~\cite{vKNM,CSB1}.  For this reason there is a
need for consistent chiral calculations for both reactions.  A first
report on the demanding $dd\to\alpha\pi^0$ calculation has been
published~\cite{CSB1}, and a calculation incorporating some of the 
physics mandated by chiral symmetry that affects
$A_{\rm fb}(np\to d\pi^0)$ exists~\cite{vKNM}, but a reliable extraction of CSB
parameters necessitates a consistent chiral calculation of the
reaction $np \to d \pi^0$.

In this paper we will use the power counting appropriate for pion
production to establish the leading- and next-to-leading-order (LO and NLO)
diagrams that contribute to the CI $s$-wave $np\to d\pi^0$ amplitude
close to threshold.  
This is a preparatory, but important, step toward the calculation of the 
CSB asymmetry, since Eq.~(\ref{eq:Afb}) makes it clear that without an 
understanding of the CI cross section the asymmetry is not under control.

The organization of the paper is as follows.  We introduce the
theoretical framework in Sec.~\ref{sec-theory}.  The LO operator
and results are presented in Sec.~\ref{sec-LO}, while the NLO operator
is introduced in Sec.~\ref{sec-NLOop}.  The results at this order are
discussed in Sec.~\ref{sec-NLO}, and the energy dependence of the
$s$-wave cross section in Sec.~\ref{sec-endep}.  We conclude
in Sec.~\ref{sec-concl}.

\section{Theory}
\label{sec-theory}

\subsection{General considerations}
\label{sec-general}

The $np\to d\pi^0$ reaction can easily be decomposed into partial waves.
The lowest CI partial waves are $^3P_1s$, $^1S_0p$, and $^1D_2p$ in the 
spectroscopic notation $^{2S+1}L_Jl_\pi$, where $S$, $L$, and $J$ are the total
spin, orbital, and total angular momentum of the incoming nucleon pair and
$l_\pi$ is the orbital angular momentum of the emerging pion.
The lowest CSB partial waves are $^1P_1s$, $^3S_1p$, $^3D_1p$, and $^3D_2p$.

As we will restrict ourselves to near-threshold CI $s$-wave pion production, 
only the $^3P_1s$ transition will be evaluated in this paper. 
This wave contributes to $\alpha$, while $^1S_0p$ and $^1D_2p$
both contribute to $\beta$.  
Note, however, that $\alpha$ and $\beta$ do not give an unambiguous 
separation of $s$- and $p$-waves.
An $s$-wave amplitude that is dependent on the pion energy, \EG, the WT
term, will produce a term quadratic in $\eta$.
This term's interference with the $\eta$-independent piece of the 
${}^3P_1s$-wave amplitude mimics the $\eta^3$ dependence of $p$-wave 
contributions to $\sigma$.  
A separation of the $\eta^2$ piece of the ${}^3P_1s$-wave amplitude and the 
$p$-wave contributions can be made using the energy dependence of the 
analyzing power $A_y$.  
Defining the calculated $s$- and $p$-wave contributions $\bar\alpha(\eta)$ 
and $\bar\beta(\eta)$, their energy dependence can be written as
\begin{eqnarray}
  \bar\alpha(\eta) & = & \bar\alpha_0+\bar\alpha'\eta^2, \nonumber \\
  \bar\beta(\eta) & = & \bar\beta_0+\bar\beta'\eta^2,
\end{eqnarray}
and the analyzing power as (again ignoring relative phases)
\begin{equation}
  A_y = \frac{\sqrt{\bar\alpha_0+\bar\alpha'\eta^2}
    \sqrt{\bar\beta_0+\bar\beta'\eta^2}\eta^2}
  {(\bar\alpha_0+(\bar\alpha'+\bar\beta_0)\eta^2)\eta} 
  \approx \sqrt{\frac{\bar\beta_0}{\bar\alpha_0}}\eta 
  \left[1+\left( \frac{\bar\beta'}{2\bar\beta_0}-\frac{\bar\alpha'}
    {2\bar\alpha_0}-\frac{\bar\beta_0}{\bar\alpha_0}\right)\eta^2\right].
\label{eq:Ay}
\end{equation}
This expansion is valid as long as 
$\frac{\bar\alpha'}{\bar\alpha_0}\eta^2\ll 1$, 
$\frac{\bar\beta'}{\bar\beta_0}\eta^2\ll 1$, and
$\frac{\bar\alpha'+\bar\beta_0}{\bar\alpha_0}\eta^2\ll1$.
(In fact, existing data for $pp\to d\pi^+$ are consistent with 
$A_y\propto\eta$ for $\eta<0.5$~\cite{ppdpi}.)

With our normalization, the total $np\to d \pi^0$ cross section is
\begin{equation}
  \sigma = \frac{\eta}{2}\frac{m_n m_p}{4\pi^2s}\frac{m_\pi}{p}
  \frac{1}{4}\sum_{\rm spins}|\mathcal{M}|^2,
\end{equation}
where $\bp$ is the relative c.m.\ momentum of the incoming nucleons, 
$\sqrt{s}$ is the total energy in the c.m.\ system, and 
the sum is over the nucleon and deuteron spins.
A general form for the $np\to d\pi^0$ matrix element, $\mathcal{M}$, in 
$p$-space, is then
\begin{equation}
  \mathcal{M} = \sqrt{2E_d}\int\frac{d^3p''d^3p'}{(2\pi)^6}
  \psi_d^\dagger(\bp'')\mathcal{A}(\tilde\bp)\psi_{np}(\bp',\bp),
\label{eq:M}
\end{equation}
where $E_d$ is the deuteron energy, $\psi_d(\bp'')$ and $\psi_{np}(\bp',\bp)$ 
are the deuteron and $np$ wave functions, $\tilde\bp=\bp''-\bp'$, and 
$\mathcal{A}(\tilde\bp)$ is the pion-production amplitude.
A recurring problem in any calculation of threshold pion-production amplitudes 
is the mismatch between initial- and final-state nucleon momenta.
Since the initial loop momentum $p'\sim p\sim P\approx360$~MeV, while the 
final loop momentum $p''\sim45$~MeV, the two integrals in the matrix element 
need to be performed in quite different momentum domains.
Critical questions are: Where is the large initial-state momentum 
absorbed in order to effect the transition between the two regimes?
Does it have to pass through the deuteron wave function or are there
pieces of the operator that can help absorb it?

\subsection{Chiral perturbation theory}
We will use the ``hybrid'' approach to derive the $\chi$PT prediction
for $np \rightarrow d \pi^0$ at threshold.  In this approach chiral
power counting is applied to the transition operator $\mathcal{A}$.
The wave functions in Eq.~(\ref{eq:M}) account for the effects of
initial- and final-state nucleons propagating nearly on their mass
shell. Such states have small energy denominators that do not obey the
usual $\chi$PT power counting~\cite{Weinberg}. Thus they have to be 
calculated non-perturbatively, ideally in a manner consistent with chiral 
symmetry.  
The hybrid approach has been applied successfully to a wide range of pionic 
and electroweak reactions in the two-nucleon system (see
Refs.~\cite{Be00,birareview,Ph05} for reviews).

We use the chiral Lagrangian as developed in Ref.~\cite{Cohen},
Eqs.~(3) and (6).  These contain pieces from the first and second
chiral orders, which is all we need for the present purposes.  
The chiral Lagrangian is used to construct the transition operator 
${\cal  A}$ in Eq.~(\ref{eq:M}), which is just the sum of all of the
two-nucleon-irreducible diagrams that contribute to $NN \rightarrow NN
\pi$.  These diagrams are arranged using a power counting which
involves a number of different energy/momentum scales. 
Ordered by size these scales are:
\begin{itemize}
\item The pion momentum $|\qi{}|$ which is $\ll m_\pi$ close to threshold.
  At NLO (which is all we consider here), this scale does not enter the
  $s$-wave production amplitude.

\item The deuteron binding momentum $\gamma\equiv\sqrt{BM}=45$~MeV,
  where $B$ is the deuteron binding energy. We expect that the
  integral over ${\bf p}''$ in Eq.~(\ref{eq:M}) will be dominated by
  $|{\bf p}''| \sim \gamma$. However, here we will conservatively
  adopt $|{\bf p}''| \sim m_\pi$. 

\item The pion mass $m_\pi$ for which we take $138.039$ MeV.

\item The Delta-isobar--nucleon mass difference 
  $\Delta=M_\Delta-M=290\ {\rm MeV}\approx2m_\pi$.

\item The nucleon momentum needed to produce a pion at threshold 
  $P\equiv\sqrt{m_\pi(M+m_\pi/4)}=360$~MeV.

\item The chiral symmetry breaking scale $\Lambda_\chi\sim1$~GeV.

\end{itemize}
We will see below that the scale $\Delta$ will never appear explicitly
in the tree-level diagrams, since it will be combined with the pion
mass in propagators. Meanwhile, in loop diagrams involving the Delta-isobar,
we follow Ref.~\cite{HK} and assign $\Delta\sim P$.

Thus we have a three-scale problem, with the hierarchy 
$m_\pi\ll P\ll\Lambda_\chi$. 
We choose to expand our amplitudes in the dimensionless parameter
\begin{equation}
  x = \frac{m_\pi}{P} =   \sqrt{\frac{m_\pi}{M}} 
\approx \frac{P}{\Lambda_\chi}  \approx 0.4.
\end{equation}

\begin{figure}[th]
  \includegraphics*{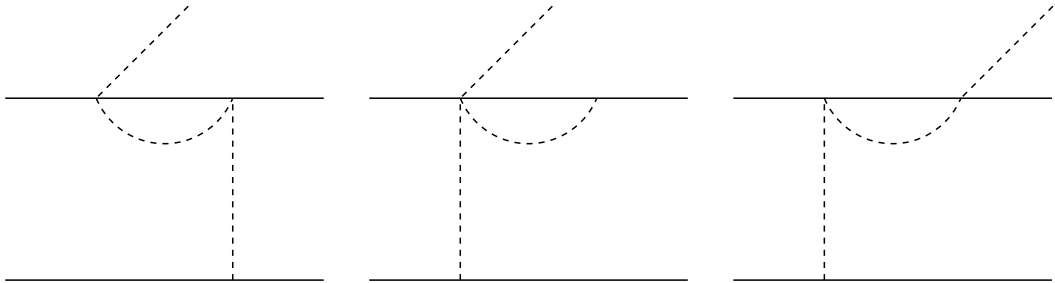}
  \caption{Examples of diagrams where the large momentum need not appear in 
    a pion loop.}
\label{fig:preduce}
\end{figure}

When counting powers of $x$ in two-nucleon-irreducible loop diagrams
it is important to distinguish between graphs of the type in
Fig.~\ref{fig:preduce} and those of the type shown in 
Fig.~\ref{fig:sNLO}. In the first case the large nucleon momentum $P$ may
pass through the nucleons and exchanged pions only (Fig.~\ref{fig:preduce}), 
whereas in the second it must go through pions in a loop [\EG,
  Fig.~\ref{fig:sNLO}(a)--(h)].
In consequence the loops in Fig.~\ref{fig:preduce} are independent of
the initial nucleon momentum $P$, since the loop resides only on a
nucleon which is moving with momentum $P$. By translation invariance,
the loop integral then scales with $m_\pi$ only.  
In Fig.~\ref{fig:sNLO} the loop integrals pick up the large momentum
scale and the integral will be dominated by pion modes of momentum
$P$.

\subsection{Wave functions consistent with chiral symmetry}

In evaluating the matrix element (\ref{eq:M}) we would like to employ
wave functions which are consistent with the chiral power counting
that we use for the operators.  
One way to do this for the deuteron was developed in Ref.~\cite{PC}.  
This method was extended to $NN$ scattering in~\cite{EFTann}.  
In these wave functions the long-range part of the $NN$ potential is 
derived using $\chi$PT~\cite{Weinberg,bira,evgeny}. 
At the same time our ignorance about the short-distance physics of the 
nucleon-nucleon interaction is parametrized by matching, at some chosen radius 
$R$, the wave function obtained for $r>R$ to a square well solution for $r<R$.
Here we will use this approach to construct the initial-state $np$ wave 
function, starting from the $^3P_1$ $np$ phase shift of the Nijmegen phase 
shift analysis~\cite{NijmPSA}.  
Since we are dealing with c.m.~energies of order $m_\pi$, we determine the 
asymptotic scattering wave functions from the Nijmegen phase shifts at a 
lab.~energy of 279.5~MeV and then expand around this point.  
This is different from the procedure chosen in~\cite{EFTann}, where, due 
to the low $NN$ energy, the $nn$ scattering length and effective range could 
be used as input.  
Note that, at the present state of the calculation, we include only
one-pion exchange (OPE) with $f_{\pi NN}^2=0.075$~\cite{NijmpiNN} as the 
long-range part of the chiral $NN$ potential.  
In the future we will also implement chiral two-pion
exchanges (TPE)~\cite{bira,evgeny,RTFdS}.  
(Unfortunately, it is impossible to use high-order chiral potentials
directly, since in general they are accurate only for energies below 
250~MeV~\cite{evgeny,EM}.)
As a cross check, below we also use some of the available low-$\chi^2$ 
potential model wave functions~\cite{NijmPot,av18} to calculate the 
matrix element (\ref{eq:M}).

The chiral wave functions we have constructed for use in our
calculation have a co-ordinate space regulator present in them: the
square well of radius $R$. Other regulators (e.g., Gaussians or sharp
cutoffs in momentum space) have been used in obtaining $NN$ wave
functions from a potential derived via a $\chi$PT
expansion~\cite{bira,evgeny,Pa99,BBSvK01,No05}. In all of these
studies a regulator is introduced before the potential is iterated
using the Schr\"odinger equation.  This means that when, in
Eq.~(\ref{eq:M}), matrix elements of pion-production operators are
computed using such wave functions, there is a
scale-dependent regularization in the ${\bf p}'$ and ${\bf p}''$
integrations.  The scale dependence of this regularization is less
obvious when $NN$ potential models are used to calculate the $\psi$'s 
appearing in Eq.~(\ref{eq:M}). However, even there it seems clear that
the hadronic form factors in such models somehow implicitly provide a
regularization that is explicitly present when wave functions derived
from the effective field theory are used.

\section{Leading-order operator and result}
\label{sec-LO}
The LO [or $\mathcal{O}(x)$] $s$-wave operator is given by the 
Weinberg-Tomozawa (WT) and the (nucleon recoil) one-body term;
\begin{eqnarray}
  \mathcal{A}_{\rm WT}(\tilde\bp) & = & \frac{3g_A\omega_q}{16f_\pi^3}
  \varepsilon^{abc}\tau_1^b\tau_2^c
  \left(\frac{\bsigma_1\cdot(\tilde\bp-\frac{\qi{}}{2})}
       {{m_\pi'}^2+(\tilde\bp-\frac{\qi{}}{2})^2}+ 
       \frac{\bsigma_2\cdot(\tilde\bp+\frac{\qi{}}{2})}
       {{m_\pi'}^2+(\tilde\bp+\frac{\qi{}}{2})^2}\right), \nonumber \\
  \mathcal{A}_{\rm 1B}(\tilde\bp) & = & \frac{-ig_A\omega_q(2\pi)^3}{4Mf_\pi}
	  \left[\tau_1^a\delta^{(3)}\left(\tilde\bp+\frac{\qi{}}{2}\right)
	    \bsigma_1\cdot\left(\bp''+\bp'-\frac{\qi{}}{2}\right)\right. 
	    \nonumber \\ & & 
	    \left.-\tau_2^a\delta^{(3)}\left(\tilde\bp-\frac{\qi{}}{2}\right)
	    \bsigma_2\cdot\left(\bp''+\bp'+\frac{\qi{}}{2}\right)\right],
\label{eq:sLO}
\end{eqnarray}
where ${m_\pi'}^2=\frac{3}{4}(m_\pi^2-q^2)$.
[In order to derive Eq.~(\ref{eq:sLO}) we have assumed equal energy
sharing between the two nucleons.]

Chiral power counting for the operator ${\cal A}$ will  be a
useful guide to the size of physical mechanisms only if different
contributions to the operator are sensitive to similar momenta in the
$NN$ wave functions used in the evaluation of the matrix element
(\ref{eq:M}).  This is clearly not the case for the operators in
Eq.~(\ref{eq:sLO}): the one-body operator ``lives'' off the
high-momentum components in the incoming state, while the two-body
nature of the pion-rescattering mechanism allows better momentum
sharing, and access to the larger low-momentum components in the $np$ and
$d$ wave functions. 
Therefore in order to estimate the impact the operator ${\cal A}_{\rm 1B}$ has
on observables we follow Ref.~\cite{Cohen} and use the Schr\"odinger equation 
for the deuteron state $|\psi_{d} \rangle$ to write
\begin{equation}
  \langle \psi_d|{\cal A}_{\rm 1B}|\psi_{np} \rangle = 
  \langle \psi_d|V G_0 {\cal A}_{\rm 1B}|\psi_{np} \rangle.
\label{eq:share}
\end{equation}
(Here $V$ is the $NN$ potential used to generate the wave functions.) 
Once Eq.~(\ref{eq:share}) is invoked momentum sharing can take place 
through $V$.  So, in order to power count the operator ${\cal A}_{\rm 1B}$, 
the pion exchange and contact interaction that make up $V$ at leading order 
in $\chi$PT~\cite{Weinberg} are included in the diagrams.
This also puts the nucleons in the ``external'' states closer to being
on-shell, and facilitates attaching wave functions that are calculated
assuming on-shell nucleons. 

\begin{figure}[ht]
  \includegraphics{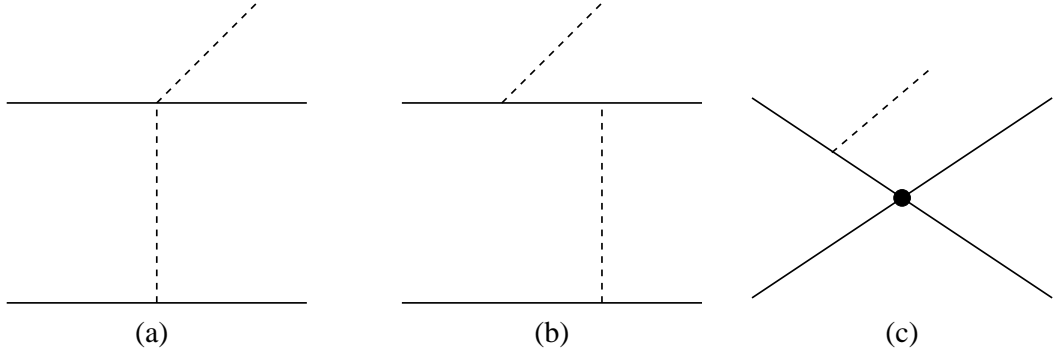}
  \caption{Leading-order diagrams for $s$-wave pion production on two 
    nucleons.}
\label{fig:sLO}
\end{figure}

The diagrams in Fig.~\ref{fig:sLO} are then the same order in chiral
power counting, which justifies the statement made at the beginning of
this section that ${\cal A}_{\rm 1B}$ and ${\cal A}_{\rm WT}$ together
give the leading-order amplitude. This statement is borne out if we
compute the different contributions to $\bar\alpha_0$ from these
leading-order processes using a variety of phenomenological and chiral
wave functions~\cite{NijmPot,av18,Bonn,PC}. The results of that
computation are shown in Table~\ref{tab:LOresults}. 
\begin{table}[ht]
  \caption{Values for $\bar\alpha_0$ obtained from various different
    contributions to the leading-order $np \rightarrow d \pi^0$
    amplitude. All results are in units of $\mu$b. Note that in each
    case (except the last column) the result for $\bar\alpha_0$ is the one 
    obtained by considering just that mechanism.}
  \begin{tabular}{c|c|cc|c|c}
    potential    &  $\bar\alpha_0$(WT) &  $\bar\alpha_0$(1B, PWIA) & 
    $\bar\alpha_0$(1B, ISI) & $\bar\alpha_0$(1B) & $\bar\alpha_0$(LO) \\ 
    \hline
    Nijm I              &  124.0 & 11.94 & 11.83 & 0.0003 & 123.6  \\
    Nijm II             &   96.8 & 16.26 & 18.21 & 0.0682 & 102.0  \\
    av18                &   82.2 & 15.75 & 22.35 & 0.6961 &  98.1  \\
    Bonn A              &  111.8 & 6.264 & 9.622 & 0.4234 & 126.0  \\
    Bonn B              &  110.4 & 11.28 & 12.96 & 0.0713 & 116.0  \\
LO $\chi$PT ($R=1.5$~fm) &  40.6 & 5.809 & 19.90 & 4.723  &  73.0  \\
LO $\chi$PT ($R=2$~fm)   & 114.7 & 1.124 & 9.391 & 4.325  & 163.6 
  \end{tabular}
\label{tab:LOresults}
\end{table}

Considering that the power counting predicts the WT and one-body term to
be of the same order, the six order of magnitude difference between the two
in the first row of Table~\ref{tab:LOresults} should cause some concern.
However, we can split up the matrix element of ${\cal A}_{\rm 1B}$ into a 
plane-wave (PWIA) and initial-state-interaction (ISI) part, \IE, write:
\begin{equation}
  \langle \psi_d|{\cal A}_{\rm 1B}|\psi_{np} \rangle = \langle
  \psi_d|{\cal A}_{\rm 1B}|{\bf p} \rangle + \langle \psi_d|{\cal
    A}_{\rm 1B} G_0 V|\psi_{np} \rangle.
\label{eq:OBme}
\end{equation}
The fact that the PWIA and ISI contributions to $\bar\alpha_0$ are of the same 
size lends credence to our application of power counting to the 
pion-production operator. 
The one-body ISI and PWIA values alone are closer to the $\bar\alpha_0$ from 
the WT mechanism than is their sum, although each piece still differs from 
$\bar\alpha_0$(WT) by an order of magnitude.
The extremely small matrix element on the left-hand side of Eq.~(\ref{eq:OBme})
apparently results from a cancellation between two numbers of a little smaller
than natural size on the right-hand side of that equation.
The total LO value for $\bar\alpha_0$ is then quite close to the number from 
Fig.~\ref{fig:sLO}(a) alone.

The numbers in Table~\ref{tab:LOresults} are in accord with those
given in the earlier study of da Rocha \EA~\cite{dRMvK}.
(Cancellations between different pieces of the integral, leading to a 
reduction in the one-body term, were also observed in Ref.~\cite{dRMvK}, 
although there the PWIA and ISI contributions are not given separately.)

We find that there is considerable wave-function dependence in the result
for $\bar\alpha_0$.
The calculation with the chiral wave function with cutoff $R=1.5$~fm is 
expected to differ from the other results, since at $p=360$~MeV the matrix 
element will probe distances where the neglected TPEs become important.
If we employ only the low-$\chi^2$ $NN$ potential models we
find a moderate spread in values: 
\begin{equation}
  \bar\alpha_0 = 111\pm13~\mu{\rm b}.
\label{eq:alphaLO}
\end{equation}  
This variation with the choice of $NN$ model reflects the
sensitivity to short-distance physics in the LO pion-production matrix element,
since these potentials reproduce the 1993 $NN$ database with 
$\chi^2 \approx 1$ and have the same OPE tail. 
The LO prediction (\ref{eq:alphaLO}) gives $\approx60\%$ of the
measured $\alpha$, which is consistent with the accuracy expected from
a LO result in an expansion in $\sqrt{m_\pi/M}$.

The use of wave functions derived assuming
zero-energy transfer between the nucleons means that there is an
inconsistency between these wave functions and the operator. 
In diagram \ref{fig:sLO}(b), for example, we assume there is energy transfer
between the nucleons, as dictated by the Feynman rules and our
assumption of equal-energy sharing between the initial particles.
If the procedure used here and in Ref.~\cite{Cohen} for
power counting diagrams by ``pulling'' the last pion exchange 
from the wave functions is accurate then this inconsistency should
be a higher-order effect, since it involves the difference
\begin{equation}
  \frac{1}{{m_\pi'}^2+(\tilde\bp\pm\frac{\qi{}}{2})^2} - 
  \frac{1}{{m_\pi}^2+(\tilde\bp\pm\frac{\qi{}}{2})^2}
  = \frac{m_\pi^2 - {m_\pi'}^2}{{m_\pi}^2+(\tilde\bp\pm\frac{\qi{}}{2})^2} 
  \times \frac{1}{{m_\pi'}^2+(\tilde\bp\pm\frac{\qi{}}{2})^2}.
\end{equation}
The momentum transferred here is of order $P$, so the difference
between including and not including energy transfer in the
one-pion-exchange propagator is of relative order
$\left(\frac{m_\pi}{P}\right)^2$, as long as $q \sim m_\pi$.  It
therefore contributes first at two orders beyond leading, which is one
order higher than we consider here.  

Similar arguments ensure that any $\eta$ dependence of the leading-order 
$s$-wave production amplitude ${\cal A}_s$ arising from energy flowing 
through these meson denominators is also a NNLO effect. 
Thus
\begin{equation}
  \mathcal{A}_s = \omega_q
  \left[\widetilde\mathcal{A}_s+\mathcal{O}(x^2\eta^2)\right],
\label{eq:alphaEdep}
\end{equation}
where $\widetilde\mathcal{A}_s$ is the reduced leading order $\mathcal{A}_S$ 
and the $\mathcal{O}(x^2\eta^2)$ correction stems from the $\qi{}$ dependence
of the vertices and propagators in Eq.~(\ref{eq:sLO}).
Thus the prediction of $\chi$PT for these diagrams (\ref{eq:alphaEdep}) is that
(ignoring ISI for the moment)
\begin{equation}
  \sigma = \frac{\eta}{2}\bar\alpha_0[1+\eta^2+\mathcal{O}(x^2\eta^2)],
\label{eq:etadep}
\end{equation}
which implies that $\bar\alpha'=\bar\alpha_0$ at LO.
However, a numerical calculation including the full scattering state
reveals that 
\begin{equation}
  \bar\alpha'=-57\pm4~\mu{\rm b}
\label{eq:alphapLO}
\end{equation}
for the potential models, and somewhat smaller, but positive 
($\bar\alpha'=41\pm30$~$\mu$b), for the chiral wave functions. 
The energy dependence coming from ISI apparently
provides a $\agt100\%$ correction to Eq.~(\ref{eq:etadep}).
We will return to this issue of energy dependence once the NLO 
contributions have been calculated.

\section{Next-to-leading order operator}

\label{sec-NLOop}

\begin{figure}[ht]
  \includegraphics{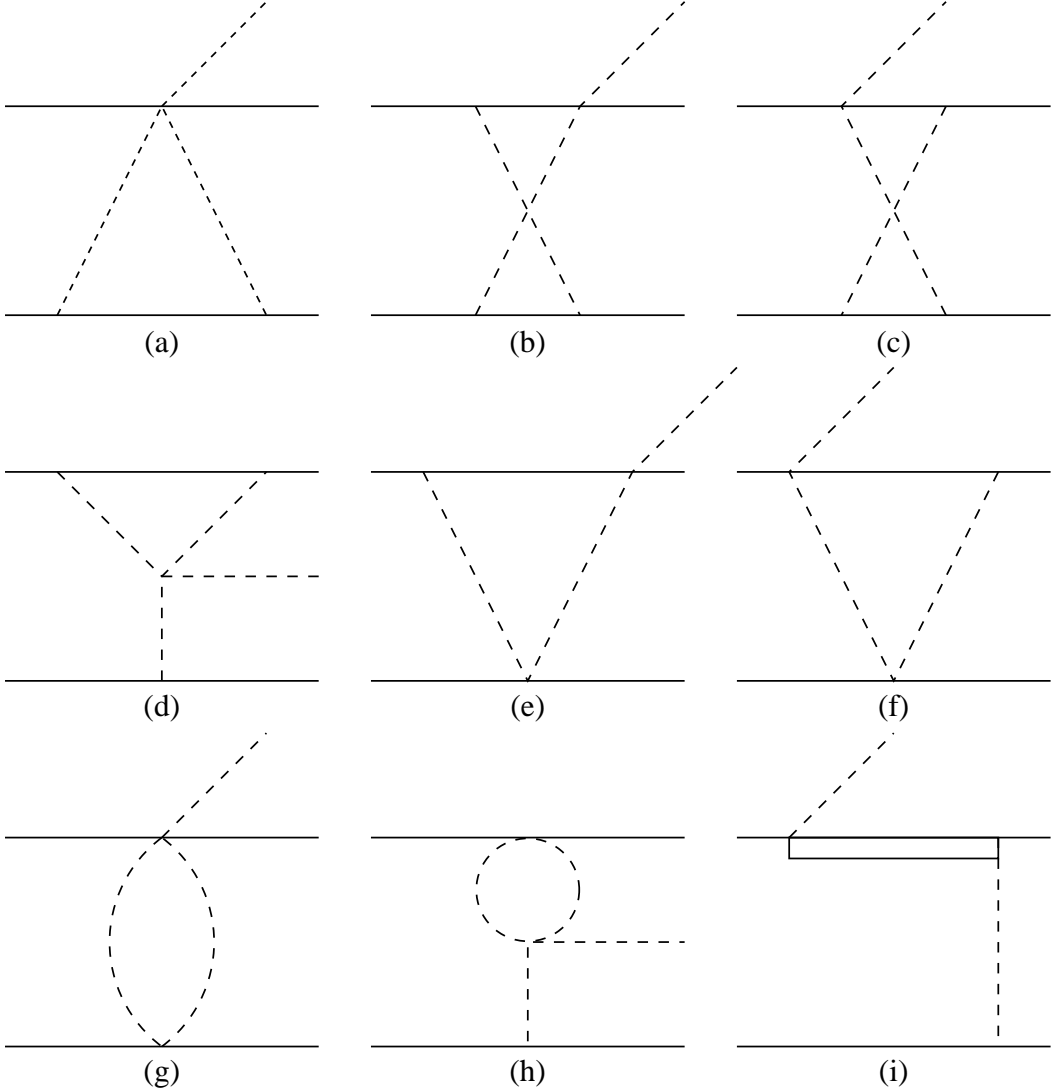}
  \caption{Next-to-leading-order diagrams (with no Delta-isobars in loops) 
    for $s$-wave pion production on two nucleons.}
\label{fig:sNLO}
\end{figure}

At NLO [or $\mathcal{O}(x^2)$], there are pion loops and a 
Delta-isobar--excitation term.  
The corresponding diagrams are shown in Fig.~\ref{fig:sNLO}. 
Some of these were discussed in Ref.~\cite{novel}, and a complete 
treatment can be found in Ref.~\cite{USC2}. 
Here we draw on the analysis of Hanhart and Kaiser~\cite{HK} for the pion 
loops shown in Fig.~\ref{fig:sNLO}, since for these graphs they employed 
the power counting in $x$ we are using here.

At NLO and in threshold kinematics diagrams \ref{fig:sNLO}(b)
and \ref{fig:sNLO}(c) cancel for $NN\to d\pi$, while \ref{fig:sNLO}(e)
and \ref{fig:sNLO}(f) $\propto m_\pi\ln m_\pi$, and thus are of higher
order~\cite{HK}.  Diagrams \ref{fig:sNLO}(g) and \ref{fig:sNLO}(h)
are also of higher order in the $x$ expansion because of their vertex
structure.  As for diagram \ref{fig:sNLO}(i), since the deuteron is isoscalar
this particular Delta-isobar diagram cannot contribute to $NN\to d\pi$.

There are also a number of Delta-isobar loop diagrams which nominally appear
at NLO, but they cancel each other in threshold kinematics.  This is
associated with the absence of a chirally-invariant counterterm at
this order~\cite{HK}.  Thus for $s$-wave pion production in $np\to
d\pi^0$, there are only two loop graphs that need to be calculated if
an answer accurate to $\mathcal{O}(x^2)$ is desired.

The sum of these two diagrams (a) and (d) can be evaluated
straightforwardly. The corresponding amplitude, together
with that for the analogous diagrams in which the roles of nucleons 1
and 2 are interchanged, gives the following result for the operator
mediating the $NN \to d \pi$ transition:
\begin{equation}
  \mathcal{A}_{\rm (a)+(d)}(\tilde\bp) = 
  \frac{1}{2}(\bsigma_1+\bsigma_2)\cdot\tilde\bp
  (i\tau_1^a-i\tau_2^a) B(-|\tilde\bp|^2),
\label{eq:qloops}
\end{equation}
with $g_A$ the axial coupling of the nucleon, $f_\pi$ the pion decay constant,
and $B$ given by the scalar integral
\begin{equation}
  B(\tilde{p}^2) \equiv \frac{g_A^3}{8 f_\pi^5} \frac{1}{i} 
  \int \frac{d^4 q}{(2\pi)^4} 
  \frac{(q+\tilde{p})\cdot q}{v \cdot q \, [(\tilde{p} + q)^2 - m_\pi^2] \,
  [q^2 -  m_\pi^2]}.
\label{eq:B}
\end{equation}
Here $v$ is the four-velocity of the heavy fermions in the
non-relativistic $\chi$PT Lagrangian. Since $v \cdot \tilde{p} \sim m_\pi$ for
threshold pion production, to the order to which we work $B$ can be
regarded as a function of $\tilde{p}^2$ alone.

If that function is computed using dimensional regularization (DR) we find:
\begin{equation}
  B_{\rm DR}(\tilde{p}^2)=\frac{g_A^3}{256 f_\pi^5} \sqrt{-\tilde{p}^2} + 
  \mathcal{O}(m_\pi).
  \label{eq:f}
\end{equation}
Clearly if $\tilde{p}^2=-M m_\pi$, as is the case in threshold
kinematics, this expression vanishes in the limit $m_\pi\to0$. This is
consistent with the chiral-symmetry requirement that the pion should
couple softly to the $NN$ system.

However, to get the result (\ref{eq:f}) the fact that the scale-less
and linearly divergent integral
\begin{equation}
  \frac{1}{i} \int \frac{d^d l}{(2 \pi)^d} \frac{1}{(v \cdot l) l^2}
\end{equation}
is zero in DR has been employed. If a scale-dependent regularization scheme
is used this integral will no longer vanish. For instance, with Pauli-Villars
regularization we find
\begin{equation}
  B_{\rm PV}(\tilde{p}^2)=\frac{g_A^3}{256 f_\pi^5}
  \left[-\frac{4 \Lambda}{\pi} +
    \sqrt{-\tilde{p}^2} + \mathcal{O}(m_\pi) +
    \mathcal{O}\left(\frac{\tilde{p}^2}{\Lambda}\right)\right],
  \label{eq:Blin}
\end{equation}
which does {\it not} vanish in threshold kinematics as the chiral limit
is taken. We will return to this point at the end of Section~\ref{sec-NLO}.

The result (\ref{eq:qloops}) and (\ref{eq:f}) agrees with that given
in Ref.~\cite{HK} if exchange diagrams are added to the
computation. Here, instead of calculating such diagrams explicitly, we
include the effects of fermion anti-symmetry by only summing over
those initial- and final-state $NN$ partial waves which are allowed by
the Pauli exclusion principle (see Section~\ref{sec-general}).  The
result (\ref{eq:qloops}) for the operator $\mathcal{A}$ is thus the
correct one to employ in Eq.~(\ref{eq:M}), since our $NN$ wave
functions are calculated on the (normalized) isospin basis.

\section{Next-to-leading order calculation}
\label{sec-NLO}
After Fourier transforming to $r$-space the matrix element
(\ref{eq:M}) becomes:
\begin{equation}
  \mathcal{M} = \sqrt{2E_d}\int d^3r\psi_d^\dagger(\ri{})\mathcal{A}(\ri{})
  \psi_{np}(\ri{},\bp),
\label{eq:MEr}
\end{equation}
where $\mathcal{A}(\ri{})=\int\frac{d^3\tilde{p}}{(2\pi)^3}
\expup{i\tilde\bp\cdot\ri{}}\mathcal{A}(\tilde\bp)$.  
The calculation of this Fourier transform for the $s$-wave tree-level diagrams 
is straightforward, but care is required when the pion loops are included.  
The Fourier transform of Eq.~(\ref{eq:qloops}) can only be carried out after 
a $p$-space regularization [\EG, by including a factor 
$\exp(-\tilde{p}^2/2 \Lambda^2)$ and then letting $\Lambda\to\infty$].
The result is
\begin{equation}
  \mathcal{A}_{\rm (a)+(d)}(\ri{}) = 
  \frac{1}{2}(\bsigma_1+\bsigma_2)\cdot\rhat
  (i\tau_1^a-i\tau_2^a)
  \frac{g_A^3}{256f_\pi^5}\frac{4}{\pi^2r^5}.
\label{eq:rloops}
\end{equation}
This highly singular operator causes a linear divergence in the matrix element
(\ref{eq:MEr}), as can be seen from the following arguments, which are valid 
for all wave functions calculated from a local $NN$ potential.
At small enough $r$ ($<R_{\rm div}$, say) the radial deuteron $s$-state wave 
function $u(r)\approx\kappa_0r$, while the radial $^3P_1$ $np$ wave function 
$v_1(r)\approx\kappa_1r^2$.
(Here the $\kappa_i$ are constants and $R_{\rm div}$ indicates where these
approximate forms no longer hold.)
Thus the short-range contribution to the matrix element behaves as
\begin{equation}
  \kappa_0\kappa_1\int_0^{R_{\rm div}} \frac{dr}{r^2}, 
\label{eq:1overr2}
\end{equation}
which is clearly divergent.  Thus the matrix element [in contrast to
  the operator Eq.~(\ref{eq:qloops})] does \emph{not} vanish in the
chiral limit, since the wave functions will still have the above
behavior as $r \rightarrow 0$ even if the pion-production threshold
becomes $E=0$ (though we would have a different value for $\kappa_1$
in that case).  The loop matrix element violates chiral symmetry! 

\begin{figure}[th]
  \includegraphics*[width=140mm]{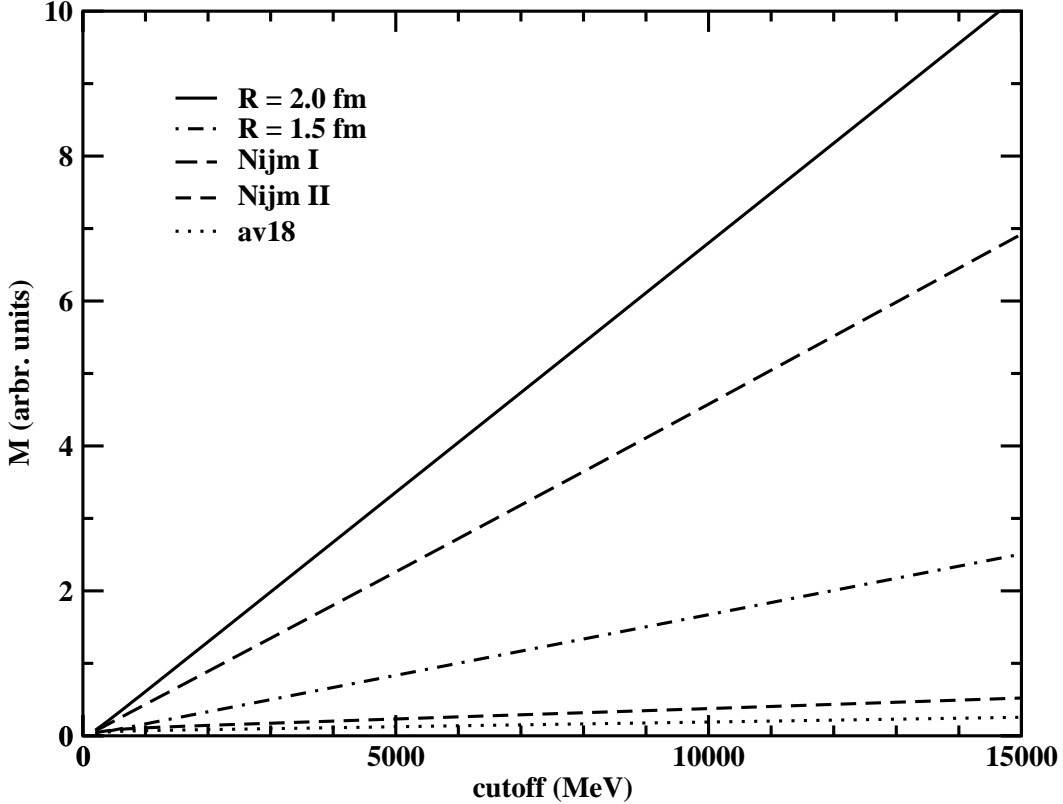}
  \caption{Cutoff ($\zeta$) dependence of the loop matrix element
    (\ref{eq:MEr}) for various wave functions as indicated. Two chiral
    wave functions with different matching radii $R$ were used, as
    well as some low-$\chi^2$ potential models.}
\label{fig:lincut}
\end{figure}

In order to render the integral (\ref{eq:1overr2}) finite we
introduce a new (coordinate-space) cutoff regulator $1-\exp(-\zeta^2r^2)$. 
We find
\begin{equation}
  \int_0^\infty dr\frac{u(r)v(r)[1-\exp(-\zeta^2r^2)]}{r^5} 
  \approx  2\kappa_0\kappa_1\zeta+C
\label{eq:lindiv}
\end{equation}
for large enough $\zeta$, since the integral is then dominated by its
behavior at small $r$.  The constant $C$ is wave-function dependent,
but independent of $\zeta$ as long as $R_{\rm div}\zeta\gg 1$.  The
linear cutoff dependence predicted in Eq.~(\ref{eq:lindiv}) is
confirmed in the actual calculations using a variety of wave functions
and cutoffs $\zeta$, see Fig.~\ref{fig:lincut}.

In the context of pionic and electromagnetic processes in the two-nucleon 
system the pion-production reactions we are considering here are special.  
In all of the other reactions that we are aware of the operators occurring 
at low order in the chiral expansion diverge much less rapidly as 
$r \rightarrow 0$ than the $\frac{1}{r^5}$ of Eq.~(\ref{eq:lindiv}). 
In consequence the matrix elements of these operators do not have the 
sensitivity to the short-distance piece of the $NN$ wave function seen in
Fig.~\ref{fig:lincut}. 
For an elegant demonstration of this in the context of pion-deuteron 
scattering, see Ref.~\cite{NH05}.

The linear divergence of Eq.~(\ref{eq:lindiv}) (or Fig.~\ref{fig:lincut}) 
is removed by introducing a short-distance operator:
\begin{equation}
  \mathcal{A}_{\rm ct}(\tilde{\bp}) \equiv  
  \frac{1}{2}(\bsigma_1+\bsigma_2)\cdot\tilde\bp
  (i\tau_1^a-i\tau_2^a)
  \frac{\tilde{d}_0(\zeta)}{Mf_\pi^3},
\label{eq:CT}
\end{equation}
where $\tilde{d}_0(\zeta)$ is a dimensionless, cutoff-dependent constant.
The operator (\ref{eq:CT}) does not come from any term present in the
original Lagrangian of~\cite{Cohen}, but such a contribution is
required here because cutoff regularization is employed to compute the
matrix element (\ref{eq:MEr}).  

The operator (\ref{eq:CT}) was not included in the analysis of Cohen
\EA~\cite{Cohen} because the pion does not couple derivatively (or via
powers of the pion mass) in $\mathcal{A}_{\rm ct}$, and therefore
this operator cannot be derived from the Lagrangian of $\chi$PT.  
The analysis of Ref.~\cite{Cohen} uses that Lagrangian, together with naive 
dimensional analysis (NDA), to order all of the pion-production operators that 
are consistent with QCD's chiral symmetry and the pattern of its breaking. 
In this case the regulator we have used when Fourier transforming the
loop amplitude (\ref{eq:qloops}) breaks the chiral symmetry invoked
when NDA was applied by Cohen \EA~\cite{Cohen}.  Therefore we expect to
have to add extra operators which break the symmetry in the same way
as the regulator. Including such terms allows us to write down a
renormalized amplitude that respects the symmetry---even though
neither the loops or counterterm individually do. The term
(\ref{eq:CT}) must be included in our calculation, in spite of the
fact that it violates chiral symmetry, so that it can cancel the
chiral-symmetry-violating divergence (\ref{eq:lindiv}) that is
proportional to the cutoff $\zeta$.

Calculating the matrix element of $\mathcal{A}_{\rm ct}$ in $r$-space, we find
[via Eq.~(\ref{eq:MEr})]
\begin{equation}
  \mathcal{M}_{\rm ct} = -\beps_d^\dagger\cdot(\phat\times\beps_{np})
  \frac{6\tilde{d}_0(\zeta)}{Mf_\pi^3}\sqrt{2E_d}\kappa_0\kappa_1,
\label{eq:ctme}
\end{equation}
where $\beps_{np}$ is the polarization vector of the spin-1 nucleon
pair and the Fourier transform creates a $\delta^{(3)}(\ri{})$ which
picks up properties of the wave functions at the origin only.  
This can be evaluated without any regularization. 
It should come as no surprise that the
expression (\ref{eq:ctme}) also does not vanish in the chiral limit.
Of course, one can always choose the value of $\tilde{d}_0$ such that
the sum of matrix elements for the loops and the counterterm vanishes
in the limit $m_\pi \rightarrow 0$.  However, since we are interested
in pions with observed masses, we choose to extract
$\tilde{d}_0(\zeta)$ at the physical threshold.  This is to say, we
fix the value of $\tilde{d}_0(\zeta)$ by fitting the experimental
value of the threshold cross section~\cite{Hutcheon}.  Given
the behavior of the loop integral, $\tilde{d}_0$ will then run
linearly with the cutoff $\zeta$.  By construction the
total matrix element will now reproduce the value of $\alpha$ from
Ref.~\cite{Hutcheon}.  Thus the renormalization not only removes the
divergence in the $np \to d \pi^0$ matrix element that occurs at NLO,
it also removes the wave-function dependence that we found in the LO 
result in Section~\ref{sec-LO}.

\begin{figure}[th]
  \includegraphics*[width=140mm]{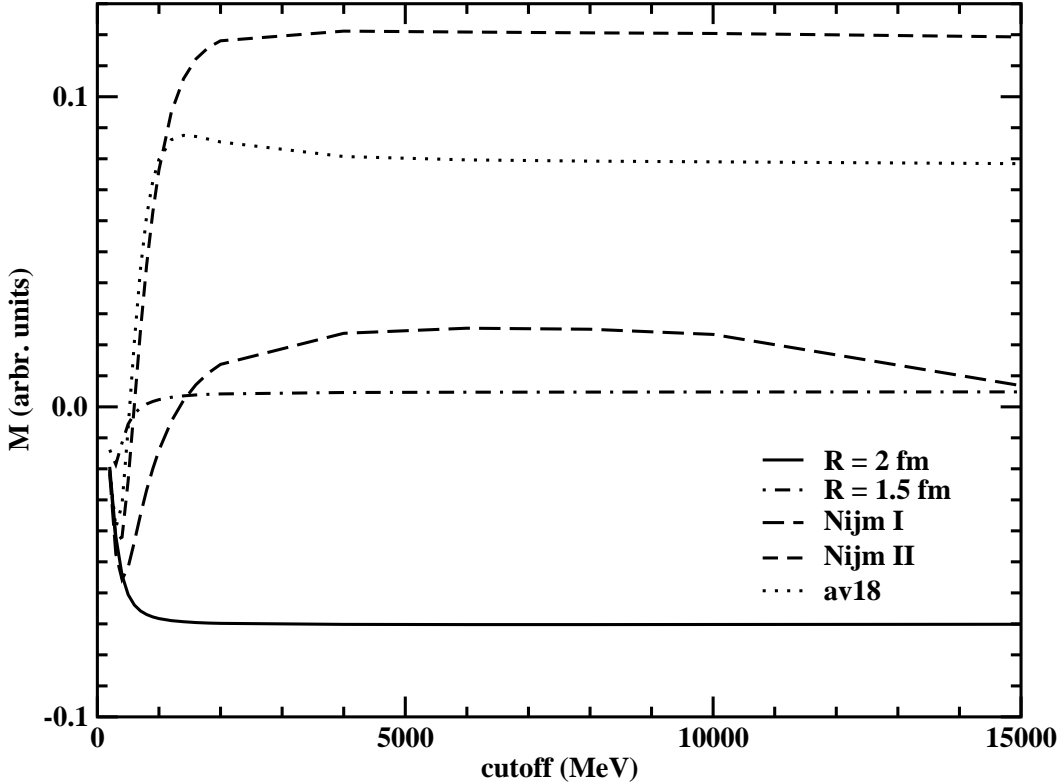}
  \caption{Single-regulator (Gaussian) cutoff ($\Lambda$) dependence of the 
    loop matrix element for various wave functions as indicated.}
\label{fig:wfdep}
\end{figure}

An alternative regularization of the matrix element of graphs
\ref{fig:sNLO}(a) and \ref{fig:sNLO}(d) is provided by
refraining from taking $\Lambda \rightarrow \infty$ directly after
Fourier transforming, and instead keeping the regulator mass finite until
after the matrix-element integral (\ref{eq:MEr}) has been performed.
If we do this with a Gaussian regulator $\exp(-\tilde{p}^2/2\Lambda^2)$ the
loop matrix element becomes
\begin{equation}
  \mathcal{M}_{\rm loops}= -\beps_d^\dagger\cdot(\phat\times\beps_{np})
  \frac{g_A^3}{32f_\pi^5}\sqrt{\frac{2E_d}{\pi^3}}\Lambda^5
  \int dr u(r)\tilde{I}(\Lambda r)v_1(r),
\label{eq:gaussianme}
\end{equation}
where 
$\tilde{I}(x)\equiv\frac12[\frac1x-x+\sqrt2(\frac{1}{x^2}+2-x^2)F(-x/\sqrt2)]$
and $F(x)[=-F(-x)]$ is the Dawson integral
\begin{equation}
  F(x) \equiv \expup{-x^2}\int_0^x dt\expup{t^2}.
\end{equation}
Surprisingly, in this procedure the coefficient of the linear
divergence vanishes for local wave functions. But, it is still the
case that the value found for the loop matrix element is sensitive to
the short-distance physics. The result is strongly dependent on the
wave function (see Fig.~\ref{fig:wfdep}).

The term linear in $\Lambda$ vanishes because there is a cancellation
between the short- and long-distance (as separated by $1/\Lambda$)
behavior of the integrand.  If a function $F_\Lambda(\tilde{p})$ is used to
regulate the Fourier transform then the divergent piece of the
integral is [for fixed $\Lambda$ and using the $r<R_{\rm div}$ limits
of $u(r)$ and $v_1(r)$]:
\begin{eqnarray}
  \lefteqn{\int_0^{R_{\rm div}} dr u(r)v_1(r) \, \frac{\partial}{\partial r}
    \int\frac{d^3\tilde{p}}{(2\pi)^3}\tilde{p}
    \expup{i\tilde\bp\cdot\ri{}}F_\Lambda(\tilde{p})}
  \nonumber \\
%
  & = & \frac{\kappa_0\kappa_1}{2\pi^2}
  \left[-r^2\frac{\partial^2}{\partial r^2}F_\Lambda(r)+
    3r\frac{\partial}{\partial r}F_\Lambda(r)-3F_\Lambda(r)\right]_{r=0}^\infty
  \nonumber \\
  && \qquad \qquad - \quad \frac{\kappa_0\kappa_1}{2\pi^2}\left[-r^2
    \frac{\partial^2}{\partial r^2}F_\Lambda(r)+
    3r\frac{\partial}{\partial r}F_\Lambda(r)-
    3F_\Lambda(r)\right]_{r=R_{\rm div}}^\infty
  \nonumber \\
  & \longrightarrow & \frac{\kappa_0\kappa_1}{2\pi^2}
  \left[-r^2\frac{\partial^2}{\partial r^2}F_\Lambda(r)+
    3r\frac{\partial}{\partial
      r}F_\Lambda(r)-3F_\Lambda(r)\right]_{r=0}^\infty
 - \quad \frac{4\kappa_0\kappa_1}{\pi^2R_{\rm div}},
\label{eq:lambdadep}
\end{eqnarray}
where we have taken the limit $\Lambda\to\infty$ in the second term on
the last line, and $F_\Lambda(r)\equiv\int_0^\infty 
d\tilde{p}\sin(\tilde{p}r)F_\Lambda(\tilde{p})$.  
Any regulator for which the square bracket in the last line vanishes 
(a class which includes exponential and Gaussian regulators) produces a 
$\Lambda$-independent result in the $\Lambda \rightarrow \infty$ limit. 
Therefore the coefficient of the linear divergence is identically zero 
for such regulators. 
And indeed, Fig.~\ref{fig:wfdep} (calculated with a Gaussian cutoff) shows 
that calculations with a variety of wave functions produce  flat
behavior for 1--2~GeV~$\alt\Lambda\alt10$~GeV.  
(Qualitatively similar curves result if an exponential cutoff is used.)

For the local potentials, the flatness region persists at least up to
$\approx200$~GeV.  
Non-local potentials (\EG, Nijm I, Bonn models) start to deviate from
flatness around 10~GeV, a hint of which is visible in Fig.~\ref{fig:wfdep}.  
Their non-locality introduces additional high-momentum components that 
influences the wave functions at small $r$.  
These high-momentum components are neglected in the two-step procedure
and there is no distinction between local and non-local potentials in 
Fig.~\ref{fig:lincut}.

The results of Fig.~\ref{fig:wfdep} confirm the prediction of
Eq.~(\ref{eq:lambdadep}). They do not show a divergence as the cutoff
mass $\Lambda$ is taken to infinity.  However, the matrix element remains
non-vanishing in the chiral limit, and its value for $m_\pi=0$ will be
determined by the high-energy mass scales $\kappa_0$, $\kappa_1$, and
$R_{\rm div}$. These mass scales are related to the (implicit or
explicit) scale-dependent regularizations carried out in obtaining the
$NN$ wave functions we have used to calculate the matrix element.  In
order to remove this dependence on the (unphysical) wave-function
regularization mass scale(s) we again introduce a short-distance
operator of the structure (\ref{eq:CT}). But this time we retain the
regulator after Fourier transforming the operator to co-ordinate
space, treating it in the same way as we did the loop-diagram operator
when we obtained (\ref{eq:gaussianme}). 
The matrix element of the short-distance operator then reads:
\begin{equation}
  \mathcal{M}_{\rm ct} = -2\beps_d^\dagger\cdot(\phat\times\beps_{np})
  \frac{\tilde{d}_0(\Lambda)}{Mf_\pi^3}\frac{\sqrt{E_d}}{\pi}\Lambda^5
  \int rdru(r)\expup{-\Lambda^2r^2/2}v_1(r).
\label{eq:CT2}
\end{equation}

We now include the short-distance operator (\ref{eq:CT2}) and adjust
$\tilde{d}_0(\Lambda)$ such that the total matrix element squared
gives $\bar\alpha_0=184$~$\mu$b. 
This yields the values for $\tilde{d}_0(\Lambda)$ shown in 
Table~\ref{tab:ctvalue}.  
The $\tilde{d}_0$'s shown there are, for most wave functions, quite rapidly 
varying with $\Lambda$. 
This is because the values given are for $\Lambda$ below the point where 
$\mathcal{M}_{\rm loop}$ becomes $\Lambda$ independent, 
see Fig.~\ref{fig:wfdep}.
\begin{table}[ht]
  \caption{Values for $\tilde{d}_0$ for various wave functions and cutoffs, 
    using ``one-step'' regularization with Gaussian.
  For each $\Lambda$ there are two solutions, labeled (1) and (2).}
  \begin{tabular}{c|cc|cc|cc}
    &  \multicolumn{2}{c|}{$\Lambda=400$~MeV} & 
    \multicolumn{2}{c|}{$\Lambda=800$~MeV} & 
    \multicolumn{2}{c}{$\Lambda=1200$~MeV} \\
    potential & $\tilde{d}_0(1)$ & $\tilde{d}_0(2)$ 
    & $\tilde{d}_0(1)$ & $\tilde{d}_0(2)$ 
    & $\tilde{d}_0(1)$ & $\tilde{d}_0(2)$ \\ \hline
    Nijm I              & 1.624 & -0.0983 & 1.557 & -0.125 & 1.669 & -0.160 \\
    Nijm II             & 1.742 & -0.199  & 2.629 & -0.482 & 4.780 & -1.079 \\
    av18                & 1.928 & -0.2536 & 4.139 & -0.857 &10.138 & -2.320 \\
    Bonn A              & 1.697 & -0.1017 & 1.944 & -0.194 & 2.365 & -0.291 \\
    Bonn B              & 1.670 & -0.1325 & 2.006 & -0.257 & 2.678 & -0.437 \\
    LO $\chi$PT(1.5 fm) & 3.331 & -0.7221 & 4.103 & -0.935 & 4.368 & -1.004 \\
    LO $\chi$PT(2 fm)   & 1.887 &  0.0072 & 1.502 & 0.0169 & 1.445 &  0.0179 
  \end{tabular}
\label{tab:ctvalue}
\end{table}

For each $\Lambda$ there are two solutions for $\tilde{d}_0$ since we are
solving a quadratic equation.
Once $p$-waves are included in the calculation it should be possible to
resolve this ambiguity with polarized data, \EG, $A_y$, 
which is available for $pp\to d\pi^+$ at low energies~\cite{ppdpi}. 
Such data fixes the relative phase between the $s$- and $p$-waves, provided 
the phase of the $p$-wave is known, see Eq.~(\ref{eq:Ay}).  
We reiterate that, regardless of which value of $\tilde{d}_0$ turns out to be 
correct, the renormalization that is necessary at NLO in the threshold 
$np \to d \pi^0$ amplitude automatically eliminates the wave-function 
dependence that is present in the LO amplitude. 

This procedure of regularizing using a Gaussian and keeping $\Lambda$
finite until after the matrix element is calculated seems more
intuitively appealing and consistent than the ``two-step'' procedure
introduced above. The fact that the term linearly proportional to the
cutoff vanishes in such a ``one-step'' regularization also means that
renormalization does not involve taking the difference of two large
numbers.  

Nevertheless, there remains a concern regarding the consistency of our
use of two different regularization schemes: DR to compute the
operators and cutoffs for the integrals we evaluate to obtain the
matrix elements of those operators.  
Ideally the same regularization should be used for both pieces of our 
calculation. 
In this context we remind the reader that introducing a Pauli-Villars 
regulator (instead of DR) in the evaluation of the diagrams \ref{fig:sNLO}(a) 
and (d) results in a linear divergence appearing in the function 
$B(\tilde{p}^2)$ [see Eq.~(\ref{eq:Blin})].  
Therefore if a scale-dependent regulator is used to compute the loops of 
Fig.~\ref{fig:sNLO}, a counterterm of the form (\ref{eq:CT}) must be 
introduced {\it already at the level of the operator}, in order to 
facilitate renormalization of this linear divergence. 
Using DR---in which the renormalization of such linear divergences is 
performed implicitly by setting scale-less, linearly divergent integrals 
to zero---to compute $B(\tilde{p}^2)$ postpones the necessity to introduce 
this counterterm. 
But it cannot be prolonged indefinitely if scale-dependent regulators are used 
to calculate the matrix element of the operator (\ref{eq:qloops}).

\section{Energy dependence of $\bar\alpha$ at NLO}
\label{sec-endep}
The NLO loops are proportional to $-\tilde{p}^2\sim M\omega_q$.
Therefore, as at LO, $\chi$PT predicts that $\bar\alpha'=\bar\alpha_0$ for 
the PWIA piece of the amplitude.
The advantage at NLO is that short-range pion-production operators are 
included, which helps ameliorate the mismatch between the initial- and 
final-state momenta. Of course, $\alpha$ is now correctly reproduced, since
the counterterm values given in Table~\ref{tab:ctvalue} are found by 
adjusting $\bar\alpha_0$ to the experimental $184$~$\mu$b.
Calculating the energy dependence of the reduced $s$-wave cross section and
fitting it to $\bar\alpha=\bar\alpha_0+\bar\alpha'\eta^2$ for various 
$\Lambda$'s, gives the values of Table~\ref{tab:alphap2}.
(For cutoffs above $\sim1$~GeV the calculation becomes independent of 
$\Lambda$.)

\begin{table}[ht]
  \caption{Values for $\bar\alpha'$ (in $\mu$b) for various potentials, 
    using counterterm values extracted for different $\Lambda$.
    The calculation is done for both $\tilde{d}_0$ solutions $(1)$ and $(2)$.}
  \begin{tabular}{c|cc|cc|cc}
    & \multicolumn{2}{c|}{$\Lambda=400$~MeV}
    & \multicolumn{2}{c|}{$\Lambda=800$~MeV} 
    & \multicolumn{2}{c}{$\Lambda=1200$~MeV} \\
    potential & $\bar\alpha'(1)$ & $\bar\alpha'(2)$ 
    & $\bar\alpha'(1)$ & $\bar\alpha'(2)$
    & $\bar\alpha'(1)$ & $\bar\alpha'(2)$ \\
    \hline
    Nijm I     &  29 & -61 &   98 & -53 &  108 & -52 \\
    Nijm II    &  54 & -78 &  144 & -62 &  169 & -59 \\
    av18       &  28 & -84 &  102 & -72 &  121 & -69 \\
    Bonn A     &  14 & -66 &   78 & -59 &   87 & -57 \\
    Bonn B     &  20 & -72 &   82 & -64 &   92 & -63 \\
    $R=1.5$~fm & 777 & 205 & 1471 & 352 & 1637 & 383 \\
    $R=2.0$~fm & 277 &  93 &  434 & 103 &  466 & 104 
  \end{tabular}
  \label{tab:alphap2}
\end{table}

Regardless of which choice we make for $\tilde{d}_0$, the chiral wave
functions give results for $\bar{\alpha}'$ that 
deviate considerably from the potential models and from each other, which 
might seem to be a disconcerting result.
However, this deviation presumably comes from our neglecting TPEs, which 
seem to play an important role in processes at this momentum transfer.
Our chiral wave functions, especially the one with $R=1.5$~fm, 
give a poor description of physics at $r<2$~fm. 
With this in mind we conclude that the range of 
NLO predictions for the two solutions is:
\begin{eqnarray}
  \bar\alpha'(1) & = & 149\pm41~\mu{\rm b},\\
  \bar\alpha'(2) & = & -60\pm8~\mu{\rm b}. \label{eq:secondsoln}
\end{eqnarray}
The fact that (in either case) $\bar\alpha'\neq\bar\alpha_0$ can be
attributed almost entirely to distortions in the initial state, since
higher-order $\qi{}$ dependence in the amplitudes is a much smaller
effect. 

In fact, using the first $\tilde{d}_0$ value yields $\bar\alpha'$s in
reasonable accord with the PWIA ansatz
$\bar\alpha'\sim\bar\alpha_0$. If, on the other hand, the second
solution is chosen, the NLO prediction for $\bar\alpha'$ has only a
very small shift from the LO result (\ref{eq:alphapLO}).  Because of
this convergence, we favor the second solution.  The chiral series for
$\bar\alpha'$ seems to be behaving well, so we anticipate that
higher-order corrections to the prediction (\ref{eq:secondsoln}) are
$\alt 15$\%. Once the $p$-waves have been calculated this 
result can be tested by comparison with, \EG, the
analyzing power.

In any case, the results of Table~\ref{tab:alphap2} indicate that some
of the experimental $\beta$ does arise from the energy dependence of
the $s$-wave pion-production amplitude. However, $p$-wave pion
production gives the greater part of the observed number.

\section{Conclusions}
\label{sec-concl}
We have calculated the $s$-wave cross section for $np\to d\pi^0$ (or,
equivalently, the Coulomb-corrected $pp\to d\pi^+$), using chiral
perturbation theory to NLO.  
We find that the dimensionally regularized NLO pion loops give a divergent 
result when they are sandwiched between wave functions, violating 
chiral-symmetry constraints. 
Thus the matrix element itself needs to be regularized.
The variation with cutoff of this matrix element depends on the choice
of regularization procedure.  Regardless of the particular procedure
chosen, a short-distance operator is needed to renormalize the theory.
Matrix elements of both the loops alone and the short-distance
operator alone are dependent on short-range physics, \IE, they display
a strong variation with choice of wave function.  However, by
construction, their sum is cutoff independent at the pion-production
threshold. 
A similar sensitivity to short-range physics has been
observed in a calculation of $pp\to pp\pi^0$~\cite{Andoetal}.

Therefore it is not immediately apparent that $\chi$PT can make more
than a leading-order prediction for the threshold amplitude in $np
\rightarrow d \pi^0$.  This prediction has a modest wave-function
dependence, but, being a LO result, is only accurate at the $\pm 40$\%
level.  However, $\chi$PT does predict the energy-dependence of the
$s$-wave pion-production operator.  With an appropriate choice of the
counterterm, this prediction is very little changed by NLO
corrections.  Pions produced in a relative $p$-wave are expected to
give the remainder of the (non-trivial) energy-dependence of the
near-threshold $np \rightarrow d\pi^0$ cross section.  Understanding
charge-independent $p$-wave pion production near threshold is also
critical to an interpretation of the TRIUMF result for $A_{\rm
  fb}(np\to d\pi^0)$, and so we are currently working on a calculation
of the amplitude for $p$-wave pion production in $np \rightarrow d
\pi^0$.

{\bf Note:} Immediately before submitting this paper we became aware of an 
ongoing study by Lensky~\EA~\cite{Le05}. They propose a solution to the 
difficulty with the NLO pion-production operator we have identified here. 
This solution involves a careful treatment of the two-nucleon-reducible and 
two-nucleon-irreducible pieces obtained when the leading-order mechanism 
depicted in Fig.~\ref{fig:sLO}(a) is sandwiched between $NN$ wave functions. 
Presumably this also leads to a different prediction for the energy dependence 
of the NLO $s$-wave $np \rightarrow d \pi^0$ cross section.

\begin{acknowledgments}
A.~G. and D.~R.~P. are grateful to Chuck Horowitz for originally
suggesting this project at the INT-03-3 program and for much
subsequent moral support and useful input.  We also appreciate
discussions with Christoph Hanhart and Fred Myhrer. We thank Lucas
Platter and Bira van Kolck for comments on the manuscript.
A.~G. thanks the Indiana University Nuclear Theory Center for
providing hospitality in the start-up phases of this work.  This work
was supported in part by the Institute for Nuclear and Particle
Physics at Ohio University and the DOE grants DE-FG02-93ER40756
(C.~E., A.~G., and D.~R.~P.) and DE-FG02-02ER41218 (D.~R.~P.).
\end{acknowledgments}

\bibliographystyle{unsrt}

\end{document}